\documentclass[10pt,letterpaper]{article}
\usepackage[top=0.85in,left=1in,footskip=0.75in,marginparwidth=2in]{geometry}

\usepackage[utf8]{inputenc}

\usepackage{cite}
\usepackage{amsmath}
\usepackage{gensymb}
\usepackage{xfrac}
\usepackage{soul}
\usepackage{xcolor}

\usepackage{nameref}
\usepackage[hidelinks]{hyperref}

\usepackage{microtype}
\DisableLigatures[f]{encoding = *, family = * }

\usepackage{changepage}

\usepackage[aboveskip=1pt,labelfont=bf,labelsep=period,singlelinecheck=off]{caption}

\makeatletter
\renewcommand{\@biblabel}[1]{\quad#1.}
\makeatother

\usepackage{lastpage,fancyhdr,graphicx}
\usepackage{epstopdf}
\fancyheadoffset[L]{2.25in}
\fancyfootoffset[L]{2.25in}

\usepackage{color}

\definecolor{Gray}{gray}{.25}

\usepackage{graphicx}

\usepackage{sidecap}

\usepackage{wrapfig}
\usepackage[pscoord]{eso-pic}
\usepackage[fulladjust]{marginnote}
\reversemarginpar

\begin{document}
\vspace*{0.35in}

\begin{center}
{\Large
\textbf\newline{Template-assisted growth of silver nanowires by electrodeposition}
}
\newline
\\
Kumaresh K R\textsuperscript{1,2},
Lakshman Neelakantan\textsuperscript{2,3\#},
Parasuraman Swaminathan\textsuperscript{1,3*}

\bigskip
\textsuperscript{1}Electronic Materials and Thin Films Lab, \\
\textsuperscript{2}Corrosion Engineering and Materials Electrochemistry Laboratory, \\
\textsuperscript{3}Ceramics Technologies Group - Center of Excellence in Materials and \\Manufacturing for Futuristic Mobility, \\
Dept. of Metallurgical and Materials Engineering, \\
Indian Institute of Technology, Madras, Chennai, India
\\

\bigskip
\#Email: nlakshman@iitm.ac.in \\
*Email: swamnthn@iitm.ac.in

\end{center}

\section*{Abstract}
Template-assisted synthesis is a facile way of growing various one-dimensional nanostructures. Among various nanoporous templates, anodic aluminium oxide (AAO) is most commonly used. In this work, the kinetics of silver nanowire growth by electrodeposition in AAO templates was investigated. The AAO templates were fabricated using the standard two-step anodisation process. The geometrical features of the fabricated templates, such as pore diameter ($55.03 \pm 6.36$ nm), interpore distance ($115.74 \pm 6.52$ nm), wall thickness ($27.78 \pm 3.4$ nm), porosity (19.87 \%), and pore density ($9.12 \times 10^9$ pores cm$^{-2}$) were obtained using scanning electron microscopy. The oxide layer at the bottom of the pores was modified by the barrier layer thinning process and chemical etching in order to facilitate uniform metal deposition. This was followed by pore widening to increase the diameter of the pores to $65.17 \pm 7.68$ nm to enhance deposition. Interpore distance increased to $119.82 \pm 5.91$ nm and wall thickness remained almost constant at $27.71 \pm 2.13$ nm. Silver was deposited using conventional electrodeposition in constant current mode. Deposition parameters such as time, current, solution concentration, and temperature were systematically investigated. For each deposition, the efficiency of the process was calculated by considering the filling fraction of the pores and the average length of the wires, all of which were measured from the SEM images. A linear growth versus time was observed within the parameter range. From the experiments, it was observed that decreasing the current, increasing the deposition temperature, and depositing for longer duration increased the average length of the nanowires. The use of electrodeposition to grow metallic nanowires can be extended to other pure metal and alloy systems. 

\bigskip

\noindent \textbf{Keywords:} Silver nanowires; Anodic aluminium oxide; Electrodeposition; Kinetics; Scanning electron microscopy;

\bigskip
\newpage
\section {Introduction}
Silver nanowires are an important part of flexible optoelectronic devices\cite{Wu2020}. They are primarily used as transparent electrodes and considered a viable replacement for indium tin oxide (ITO)\cite{Azani2020}, especially for flexible applications. Silver has one of the highest bulk conductivities among metals, while the incomplete surface coverage (arising out of the percolation network of nanowires in the deposit) makes the film optically transparent\cite{Zhou2021, Sorel2012}. Silver also possess good ductility, unlike ITO which is brittle (ceramic)\cite{Nair2019}. The aspect ratio of silver nanowires is an important parameter to gauge its efficiency as a transparent electrode. Shorter wires, i.e., smaller aspect ratios, will reduce the transparency since a greater surface coverage is needed to obtain good electrical conduction\cite{Sorel2012, Mutiso2013}. On the other hand, longer wires are difficult to deposit, especially through techniques such as extrusion or inkjet printing\cite{Nair2019}. Also, since the number of individual junctions are small, surface adhesion issues can also arise in these long wires. Silver nanowires in both pure and composite form are used extensively\cite{Nair2019, Nair2020, Nair2021, Nair2021FPE} for flexible optoelectronic devices and hence researchers are devising new/general methods of synthesis and deposition.     

The most commonly used technique for the synthesis of silver nanowires is the polyol process\cite{Sun2003, Ding2016, Zhang2017, Chen2017, Patil2021}. This is a chemical reduction process, which involves the reduction of silver from its salt (typically silver nitrate or acetate) in the presence of a reducing agent. The same process is also used to synthesize silver nanoparticles\cite{Kosmala2011}. A surfactant, usually polyvinyl alcohol (PVA), is also used to prevent the agglomeration of the individual nanoparticles and wires. By properly controlling the concentration of the surfactants it is possible produce directed growth, leading to the preferential formation of nanowires. Various growth parameters, including concentration, temperature, and pH need to be controlled in order to obtain nanowires\cite{Ding2016, Zhang2017}. The process is usually specific to silver and cannot be directly extended to other noble metals, such as copper\cite{Nam2016} or gold\cite{Halder2007, Chirea2011}, which are also used in flexible electronics. Other techniques are also available for the growth of silver nanowires, such as thermal evaporation\cite{Chen2009}. However, this process is not controllable and hard to scale up for commercial production. Template-assisted techniques are also available for nanowire production\cite{Kline2006, Liu2013}. Silver nanowires have been synthesized using polymer templates\cite{Jiang2001, Lazzara2009}. Similarly, mesoporous silica has also been used to grow silver nanowires\cite{Huang2000}. Among the various template-assisted routes, anodic aluminium oxide (AAO) is the most widely used route to fabricate one-dimensional nanostructures. 

Aluminium anodization is a popular technique to form a protective oxide film on the surface. The two-step anodization process to produce highly ordered hexagonal nanoporous alumina was reported in 1995 by Masuda and co-workers\cite{Masuda1995}. After that, a variety of researchers have studied the effect of anodization parameters such as voltage, temperature, electrolyte, and time on the growth of AAO\cite{Zhang2004, Li2008}. AAO membranes, with their highly ordered nanopore arrangement, typically serve as ideal templates for the formation of various nanostructured materials\cite{Mijangos2016, Xie2016}. The method offers good control over shape, size, and length of the wires. To deposit the wires, various processes exist, including DC electrodeposition\cite{Yin2001}, pulsed electrodeposition\cite{Kim2006}, electroless deposition\cite{Wang2006} and, photo deposition\cite{Zhao2003}. Recently, we also demonstrated the growth of copper nanowires in a AAO template using galvanic displacement\cite{Ganapathi2019}. In this work, electrodeposition is used to grow silver nanowires using a home-grown AAO template. The effect of the deposition parameters on the length of the nanowires have been studied and the process efficiency has also been calculated. These results show that electrodeposition in AAO templates offers a facile route to synthesize metallic nanowires. 

\section {Materials and methods}

AAO templates were grown using the two-step anodization process, using the procedure described in detail elsewhere\cite{Ganapathi2019}. To describe the process briefly, strips of 99.99\% pure aluminium were cut from a foil and then cleaned in acetone, ethanol, and finally, deionized (DI) water. The foils were then electropolished in a solution consisting of ethanol and perchloric acid (HClO$_4$). Electropolishing was carried out at 10 \degree C at a potential of 20 V for 60 s. Two-step anodization was then carried out on these electropolished foils. In the first step, anodization was carried out in 0.3 M oxalic acid at a voltage of 45 V. The electropolished template was the anode and a Pt gauze was used as the inert cathode. Anodization was carried out for an hour at 20 \degree C. After the first anodization, chemical etching was carried out to etch away the pores. The etchant solution used was a mixture of phosphoric and chromic acid and the process was carried out at an elevated temperature of 60 \degree C. Second anodization was carried out using the same conditions as the first anodization, but for a duration of 2 h. The process produced mostly defect-free pores which were parallel to each other. The new pores were grown from the nucleation sites created from the first anodization and this helped achieve a more uniform distribution of pores. Second anodization can be done for a longer duration too, and the thickness of the AAO layer increases linearly with time. 

The as-produced nanoporous alumina channels produced by two-step anodization are separated from the aluminium substrate by an oxide barrier layer\cite{Masuda1995}. Hence, for use of these channels for electrodeposition the barrier layer has to be removed, which is carried out by a process called Barrier Layer Thinning (BLT), developed by Furneaux \textit{et al.}\cite{Furneaux1989}. BLT leads to the formation of branched cracks and openings at the template bottom, exposing the underlying aluminium substrate. The process was immediately carried out after second anodization and it involved the reduction of the applied voltage in a regulated fashion to reduce the thickness of the barrier layer. From 45 V, the voltage was brought down at a rate of 2 V min$^{-1}$ till 29 V and then at a rate of 1 V min$^{-1}$ till 5 V. During the potential drop, the amount of electro-generated Al$^{3+}$ ions reduces and leads to lower oxide formation rate. On the other hand, at the bottom of the pores, the micro-environments are still the same and it leads to a higher dissolution rate and the barrier layer thins. Post BLT, chemical pore widening was used to remove the barrier layer completely. The process also causes a slight widening of the pores. Pore widening was carried out in phosphoric acid at 20 \degree C for 45 min. The final prepared template is now ready for nanowire deposition.  

In this work, electrodeposition was carried out to deposit silver nanowires inside the pores of the AAO template. The electrolyte prepared was a 0.05 M silver nitrate solution, prepared by mixing silver nitrate and boric acid in DI water. The contents are mixed using a magnetic stirrer for around 20 min. The template was connected to the negative terminal and a platinum mesh to the positive terminal. The source meter was adjusted to provide 5 mA constant current, with a compliance of 20 V. To investigate the effect of the deposition parameters on the length of the nanowires the following parameters were varied: deposition time, applied current, solution concentration, and bath temperature. 

Post deposition, all samples were imaged by scanning electron microscopy (SEM) in a FEI Quanta 400F instrument. Plan-view images were used to analyse the pore dimensions, while cross sectional imaging was used to measure the length of the deposited nanowires. Multiple images were recorded and analyzed to obtain reliable statistics on the average nanowire lengths. Energy dispersive spectroscopy (EDS) was used for qualitative elemental analysis.   

\section {Results and Discussion}
\subsection{AAO template characterization}
\begin{figure}[h]
    \centering
    \includegraphics[width=14cm]{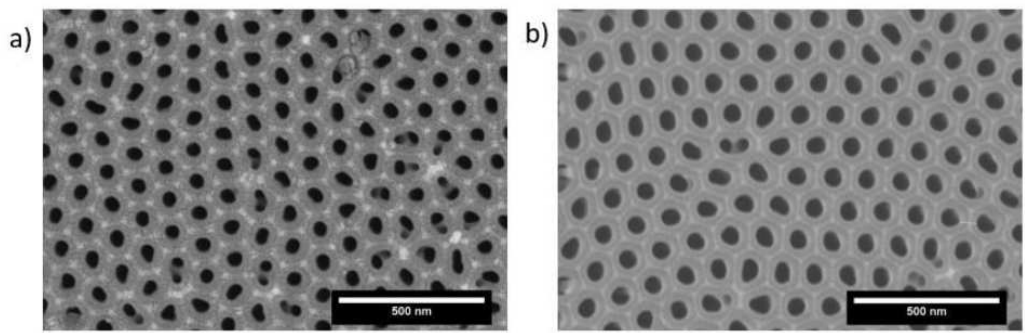}
    \caption{Representative SEM images of the AAO template (a) before and (b) after pore widening. The dimensions of the template have been extracted by analyzing similar SEM images of the templates and are tabulated in table \ref{tab:Table1}. }
    \label{fig:Fig1}
\end{figure}

In order to characterize the AAO templates SEM was carried out, before and after pore widening. From an analysis of the images, the dimensions of the AAO template were extracted. Representative SEM images of the AAO template, before and after pore widening, are shown in figure \ref{fig:Fig1}. The extracted dimensions are tabulated in table \ref{tab:Table1}.    

\begin{table}[h]
    \centering
    \caption{AAO template dimensions extracted from the SEM images, such as those displayed in figure \ref{fig:Fig1}. Pore widening causes an increase in pore diameter and porosity, while the pore areal density remains the same.  \\}
    \begin{tabular}{|c|c|c|}
    \hline
    \textbf{Parameter}&\textbf{After second anodization} & \textbf{After pore widening}  \\
    \hline
Pore diameter (nm) & 55.03 $\pm$ 6.36 & 65.17 $\pm$ 7.68 \\
Wall thickness (nm) & 27.78 $\pm$ 3.00 & 27.71 $\pm$ 2.13 \\
Interpore distance (nm) & 115.74 $\pm$ 6.52 & 119.82 $\pm$ 5.91 \\
Barrier layer thickness (nm) & 48 $\pm$ 3 & 48 $\pm$ 3 \\
Porosity (\%) & 19.87 & 26.36 \\
Pore density (cm$^{-2}$) & $9.12 \times 10^9$ & $9.12 \times 10^9$ \\
         \hline
    \end{tabular}
    \label{tab:Table1}
\end{table}

The dimensions of the AAO template can be controlled by the anodization potential, while the anodization time affects the overall length of the pores. In these experiments the length of pores is approximately 30 $\mu$m for a second anodization time of 2 h.  

\subsection{Kinetics of nanowire growth}
\begin{figure}[h]
    \centering
    \includegraphics[width=14cm]{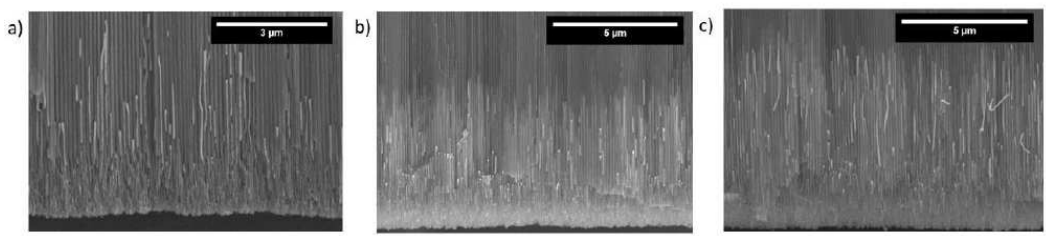}
    \caption{SEM images of the silver nanowires within AAO template for different growth times, (a) 20, (b) 40, and (c) 60 min. The overall nanowire length increases with deposition time.}
    \label{fig:Fig2}
\end{figure}

\begin{figure}[h]
    \centering
    \includegraphics[width=8.3cm]{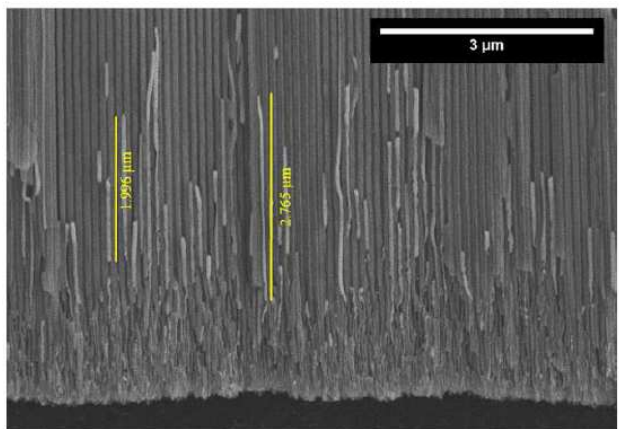}
    \caption{Two representative measurements from the 20 min deposition sample (15k magnification). The yellow lines drawn next to the wires indicate their lengths. Similar measurements were carried out for other images recorded for other deposition times.}
    \label{fig:Fig3}
\end{figure}

\begin{figure}[h]
    \centering
    \includegraphics[width=14cm]{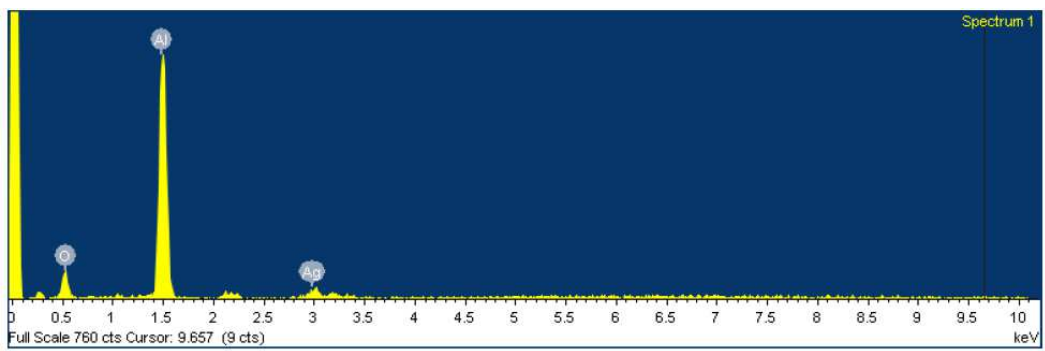}
    \caption{EDS data from the silver nanowires within the AAO template. The Al signal is from the nanoporous template, while the Ag is from the nanowires.}
    \label{fig:Fig4}
\end{figure}

Initial experiments were carried out to determine the effect of growth time on the length of the electrodeposited wires. SEM images of the wires inside the pores were taken for samples of different deposition times and representative images for three different times are shown in figure \ref{fig:Fig2}. The individual wire lengths are calculated using the ImageJ software, and an average length is calculated for a particular deposition time. The SEM image of calculated wire lengths are shown in figure \ref{fig:Fig3}. In order to confirm deposition of silver, energy dispersive spectroscopy (EDS) was carried out in the SEM and it is shown in figure \ref{fig:Fig4}.   

\begin{figure}[h!]
    \centering
    \includegraphics[width=8.3cm]{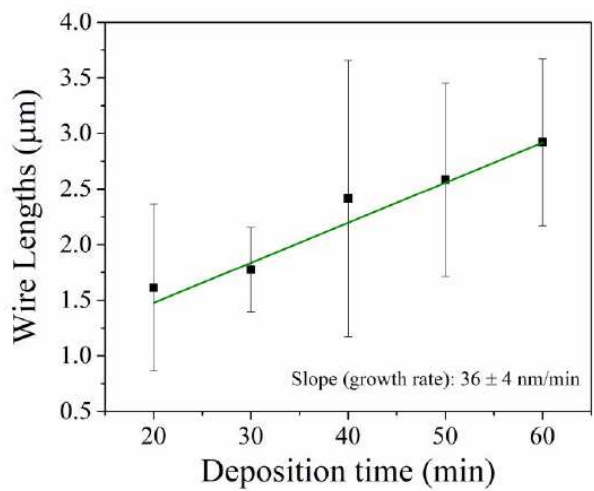}
    \caption{Plot of the nanowire length vs. deposition time. The plot is linear indicating that growth is reaction-controlled rather than diffusion-controlled. }
    \label{fig:Fig5}
\end{figure}

From the SEM images, nanowires lengths are extracted and the data is shown in \ref{fig:Fig5}. The data indicates a reaction-controlled growth rather than a diffusion-controlled growth. The linear rate law corresponds to a surface reaction being rate-limiting, the surface reaction possibly being the adsorption reaction of hydrogen on the metal surface.   

\subsection{Efficiency of the deposition process}
It is important to estimate the efficiency of the electrodeposition process since this can be used to calculate the amount of silver required for synthesis of the nanowires. Two techniques have been used in this work to estimate efficiency. One technique is based on the analysis of the SEM images to calculate the lengths of the silver nanowires and from that estimate the total volume of silver in a given area. Let $L_{av}$ denote the average length of the silver nanowires and $L_{tot}$ be the total length of the nanowires. Both can be measured from the SEM images. From the images, it is also possible to find out the fraction of pores ($f$) that are filled. Under these conditions, the ratio $f*(\sfrac{L_{av}}{L_{tot}})$ is a measure of the efficiency of the process. The efficiency values calculated by this technique are listed in table \ref{tab:Table2}.   

\begin{table}[h]
    \centering
    \caption{Efficiency calculations using the image analysis method. With this method of calculation, an increasing trend in efficiency is seen because the average length of the nanowires increases with time.  \\}
    \begin{tabular}{|c|c|c|c|}
    \hline
    \textbf{Deposition time (min)}&\textbf{$L_{av}$ ($\mu$m)} & \textbf{$f$} & \textbf{$f*(\sfrac{L_{av}}{L_{tot}})$}  \\
    \hline
20 & 1.61 & 0.57 & 0.027\\
30 & 1.77 & 0.54 & 0.028\\
40 & 2.41 & 0.55 & 0.039\\
50 & 2.59 & 0.54 & 0.042\\
60 & 2.92 & 0.55 & 0.048\\
         \hline
    \end{tabular}
    \label{tab:Table2}
\end{table}

From the table, we can conclude that the filling fraction is independent of the time of deposition. Filling fraction is an indication of the nucleation of the wires in the pores, and this
happens wherever there is a seed layer (substrate aluminium in this case). If $f$ does not depend on the deposition time, it means that the initial nucleation is constant for a given electrolyte concentration. One way to increase $f$ is to use a deposited seed layer such as Cu or Ni so that silver nucleation happens in more of the pores. Similarly, a different electrolyte can also help in increasing $f$. It should be noted that the value of $f$ tabulated in table \ref{tab:Table2} is only a lower limit, since there could be a number of pores where nucleation has occurred but not growth (and hence, they could not be counted). Since $f$ is nearly the same in all cases and $L_{avg}$ increases with time, the efficiency
also increases in the same fashion. 

A more accurate and involved method of calculating efficiency is to consider the electrochemical reactions in the system. This method involves comparing the actual mass of silver deposited and the theoretical mass of silver that should have been deposited from the electrolyte, in that time. To arrive at this efficiency, SEM images were used to obtain values of interpore distance, pore diameter, and other structural features. The volume of a single pore can be found out using the pore diameter and the length of the pore, obtained from the cross sectional view of the template, i.e. $V_1\;=\;\pi r^2 L_{tot}$. The product of the pore areal density and the area under consideration in the SEM images gives the total number of pores in that area ($n$). The product, $V_1 \times n$, provides the total volume available for silver deposition. The volume of silver actually deposited can be found using pore diameter and the average length (for a particular deposition time) i.e. $V_2\; =\; \pi r^2  L_{av}  n  f$, where $f$ is the filling fraction for that particular deposition time as described earlier. The product of $V_2$ with the mass density of silver (10.49 gcm$^{-3}$) gives the total mass of silver deposited in that particular area. Given that the current is fixed during deposition, its product with deposition time gives the total charge density. Using Faraday’s law and the molar mass of silver (107.86 gmol$^{-1}$), the theoretical mass of silver that should be deposited can be calculated. The relationship is given by
\begin{equation}
    m = \dfrac{QM}{Fz}
\end{equation}
where $m$ is the theoretical mass deposited, $Q$ is the total charge, $M$ is the molar mass of silver, $F$ is Faraday constant (96485 Cmol$^{-1}$) and $z$ is the valency (1 for silver). The ratio of the theoretical mass that can be deposited to the actual mass deposited is then the efficiency of the process. The values generated are tabulated in table \ref{tab:Table3}.

\begin{table}[h]
    \centering
    \caption{Efficiency values using the electrochemistry method. The mass of silver deposited shows an increasing trend while efficiency shows a decreasing trend. The data is in contrast to that presented in table \ref{tab:Table2} which shows an increasing efficiency with deposition time. This is because these calculations take into account the effect of deposition time on the efficiency of the process. \\}
    \begin{tabular}{|c|c|c|c|c|c|}
    \hline
    \textbf{Deposition}& \textbf{Volume of} & \textbf{Charge} & \textbf{Mass of silver}  & \textbf{Theoretical} & \textbf{Efficiency} \\
    \textbf{time}& \textbf{silver deposited} & \textbf{supplied} & \textbf{deposited}  & \textbf{mass} & \\
    \textbf{(min)}& \textbf{($\times 10^{-5}$ cm$^3$)} & \textbf{(C)} & \textbf{(mg)}  & \textbf{(mg)} & \\
    \hline
20 & 1.39 & 6 & 0.14 & 7 & 0.02 \\
30 & 1.45 & 9 & 0.15 & 10 & 0.015 \\
40 & 2.01 & 12 & 0.21 & 13 & 0.016 \\
50 & 2.12 & 15 & 0.22 & 17 & 0.013 \\
60 & 2.44 & 18 & 0.25 & 20 & 0.012\\
         \hline
    \end{tabular}
    \label{tab:Table3}
\end{table}

The data from table \ref{tab:Table3} shows a clear reverse to the trend obtained in table \ref{tab:Table2}. The method outlined here is more accurate since it is based on using Faraday's law as the benchmark in calculating the efficiency. The data also shows that the efficiency decreases with increase in deposition time. One reason for this could be a saturation limit of deposition as time increases though the plot in figure \ref{fig:Fig5} seems to indicate a linear growth regime (though the error bars indicate large variations in the individual pores). More importantly, there are competing reactions to silver reduction during deposition. Hydrogen evolution could draw current away from silver deposition leading to a lowering in efficiency with time.   

\subsection{Effect of deposition parameters}

\begin{figure}[h!]
    \centering
    \includegraphics[width=8.3cm]{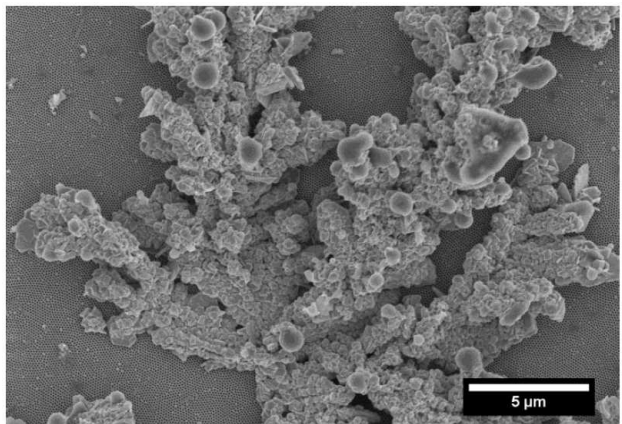}
    \caption{Silver deposited on top of the pores for 0.1 M silver concentration. This non-uniform deposition is not suitable for nanowire deposition and hence wire growth is favoured by low silver concentration. }
    \label{fig:Fig6}
\end{figure}

Various deposition parameters such as current, solution concentration, temperature, and duration (at low current) were also investigated and the final structures and efficiencies were evaluated. The normal current used during deposition was 5 mA. Two other current values, 8 and 3 mA, were evaluated. With larger current there was increased agitation in the bath due to competing reactions such as hydrogen evolution. This caused the wires to break up and hence the overall length and efficiency was lower. Lower current, 3 mA, was better in terms of wire length and efficiency though the fill factor was slightly lowered, indicating that nucleation was probably affected. When the concentration of silver in the solution was increased, deposition happened at the openings, as seen in figure \ref{fig:Fig6}. Little deposition is seen in the pores due to pore blocking but there is significant clustering at the top.  

Increasing the deposition temperature to 40 \degree C increases the length and the efficiency has also slightly increased. This can be attributed to the increased kinetics at elevated temperature. The slight increase in fill factor can also be due to the improved nucleation in the pores. A combination of two factors, long deposition time (2 h) and low current (3 mA), was also investigated. While there was some increase in length the overall efficiency was still the same as that observed earlier. 

\newpage

\section{Conclusion}

Home-grown templates were made using the two-step anodization process, followed by BLT and pore widening. The template dimensions were uniform and very close to the theoretical values. 
Electrodeposition was used to deposit silver inside the pores. EDS confirmed the presence of silver deposition and SEM images were used, along with ImageJ software, to analyse the images and lengths of the wires. The variation in the length of the wires and the filling fraction (and hence efficiency) of the process was studied by modifying different experiment conditions. Increasing deposition time increases the average length of the wires but filling fraction remains nearly the same. Similarly, increasing the current leads to smaller wire lengths but greater filling fraction and decreasing the current leads to the opposite trend. Increasing the concentration of silver in the solution hinders deposition because the aluminium layer at the bottom is not as conductive as the silver deposited on top. Increasing the deposition temperature leads to longer wires and slightly higher filling fraction. As future work, the effect of silver complex such as silver thiocyanate (AgSCN) can be studied. Silver is released more easily from complexes because of the lower
activation energy. Similarly, the solution can also be modified using additives such as suppressors, brighteners, or hydrogen poisons. These can arrest the competing reaction of hydrogen evolution and can enhance deposition of silver. 

\section*{Acknowledgments}
Support from the Ceramics Technologies Group - Center of Excellence in Materials and Manufacturing for Futuristic Mobility (project number SB/2021/0850/MM/MHRD/008275) is acknowledged. SEM imaging was carried out at the facility available in the Dept. of Chemical Engineering, IIT Madras.  

\newpage


\bibliography{library}

\bibliographystyle{unsrt}

\end{document}